\documentclass[letter]{aa}
\usepackage{graphicx}
\usepackage{natbib}
\bibpunct{(}{)}{;}{a}{}{,} 

\begin{document}

\title{Jet-disturbed molecular gas near the Seyfert 2 nucleus in M51}
\titlerunning{Jet-disturbed molecular gas near the Sy 2 nucleus in M51}

\author{S.~Matsushita \and S.~Muller \and J.~Lim}
\authorrunning{Matsushita et al.}
\offprints{S.~Matsushita, \email{satoki@asiaa.sinica.edu.tw}}

\institute{Academia Sinica, Institute of Astronomy and Astrophysics,
	P.O.~Box 23-141, Taipei 106, Taiwan, R.O.C.}


\abstract
{Previous molecular gas observations at arcsecond-scale resolution of
 the Seyfert 2 galaxy M51 suggest the presence of a dense
 circumnuclear rotating disk, which may be the reservoir for fueling
 the active nucleus and obscures it from direct view in the optical.
 However, our recent interferometric CO(3-2) observations show a hint
 of a velocity gradient perpendicular to the rotating disk, which
 suggests a more complex structure than previously thought.
}
{To image the putative circumnuclear molecular gas disk at
 sub-arcsecond resolution to better understand both the spatial
 distribution and kinematics of the molecular gas.}
{We carried out CO(2-1) and CO(1-0) line observations of the
 nuclear region of M51 with the new A configuration of the
 IRAM Plateau de Bure Interferometer, yielding a spatial resolution
 lower than 15~pc.}
{The high resolution images show no clear evidence of a disk, aligned
 nearly east-west and perpendicular to the radio jet axis, as
 suggested by previous observations, but show two separate features
 located on the eastern and western sides of the nucleus.
 The western feature shows an elongated structure along the jet and
 a good velocity correspondence with optical emission lines associated
 with the jet, suggesting that this feature is a jet-entrained gas.
 The eastern feature is elongated nearly east-west ending around the
 nucleus.
 A velocity gradient appears in the same direction with increasingly
 blueshifted velocities near the nucleus.
 This velocity gradient is in the opposite sense of that previously
 inferred for the putative circumnuclear disk.
 Possible explanations for the observed molecular gas distribution
 and kinematics are that a rotating gas disk disturbed by the jet,
 gas streaming toward the nucleus, or a ring with another smaller
 counter- or Keplarian-rotating gas disk inside.}
{}


\maketitle

\section{Introduction}
\label{sect-intro}

Active Galactic Nuclei (AGNs) are believed to be powered by gas
accretion.
This gas is supplied from interstellar matter in host galaxies,
and the gas may form rotationally-supported structures around the
central supermassive black hole.
If they are viewed close to edge-on, they may obscure the central
activity from direct view.
AGNs can be categorized as type 1 if seen face-on, and type 2 if seen
edge-on; this explanation is known as a unified model
\citep[e.g.][]{ant85}.
Indeed, a few hundred pc resolution molecular gas imaging toward
the central regions of the Seyfert 2 galaxies NGC 1068
\citep{pla91,jac93} and M51 \citep{koh96} show strong peaks
at the nuclei with velocity gradients perpendicular to radio jets,
which suggest the existence of edge-on circumnuclear rotating disks.
Recent $\sim50$~pc resolution imaging studies toward
NGC 1068 and the Seyfert 1 galaxy NGC 3227 support
this view, showing more detailed structures, namely warped disks
\citep{sch00a,sch00b}.
However, observations toward a few low activity AGN galaxies with
$<100$~pc resolution show lopsided, weak, or no molecular gas
emission toward the nuclei \citep[e.g.][]{gar03,gar05}.

M51 (NGC 5194) has also been observed in detail
with molecular lines in the past, since it is one of the nearest
\citep[7.1~Mpc;][]{tak06} Seyfert galaxies.
A pair of radio jets emanates from the nucleus and narrow line
regions (NLRs) are associated with the jet
\citep[e.g.,][]{cra92,gri97,bra04}.
Interferometric images in molecular gas show blueshifted emission on
the eastern side of the Seyfert 2 nucleus, and redshifted gas on the
western side \citep{koh96,sco98}.
This shift is almost perpendicular to the jet axis, and the estimated
column density is consistent with that estimated from X-ray
absorption toward the nucleus, suggesting that the molecular gas can
be a rotating disk and play an important role in obscuring the AGN.
Interferometric CO(3-2) observations suggest a velocity gradient
along the jet in addition to that perpendicular to the jet
\citep{mat04}.
These results imply more complicated features than a simple disk
structure.
We therefore performed sub-arcsecond resolution CO(2-1) and CO(1-0)
imaging observations of the center of M51 to study the
distribution and kinematics of the molecular gas around the AGN in
more detail.

\section{Observation and data reduction}
\label{sect-obs}

We observed CO(2-1) and CO(1-0) simultaneously toward the nuclear
region of M51 using the IRAM Plateau de Bure Interferometer.
The array was in the new A configuration, whose maximum baseline
length extends to 760~m.
Observations were carried out on February 4th, 2006.
The system temperatures in DSB at 1~mm were in the range 200-700~K,
except for Antenna 6, for which a new generation receiver gave system
temperatures of 150-230~K.
Those in SSB at 3~mm were in the range 140-250~K for Antenna 6, and
220-550~K for other antennas.
Four of the correlators were configured to cover a 209~MHz
(272~km s$^{-1}$) bandwidth for the CO(2-1) line, and a 139~MHz
(362~km s$^{-1}$) bandwidth for the CO(1-0) line.
The remaining four units of the correlator were configured to cover a
550~MHz bandwidth for continuum observations and calibration.
The strong quasar 0923+392 was used for the bandpass
calibration, and the quasars 1150+497 and 1418+546
were used for the phase and amplitude calibrations.

\begin{figure}
\resizebox{\hsize}{!}{\includegraphics{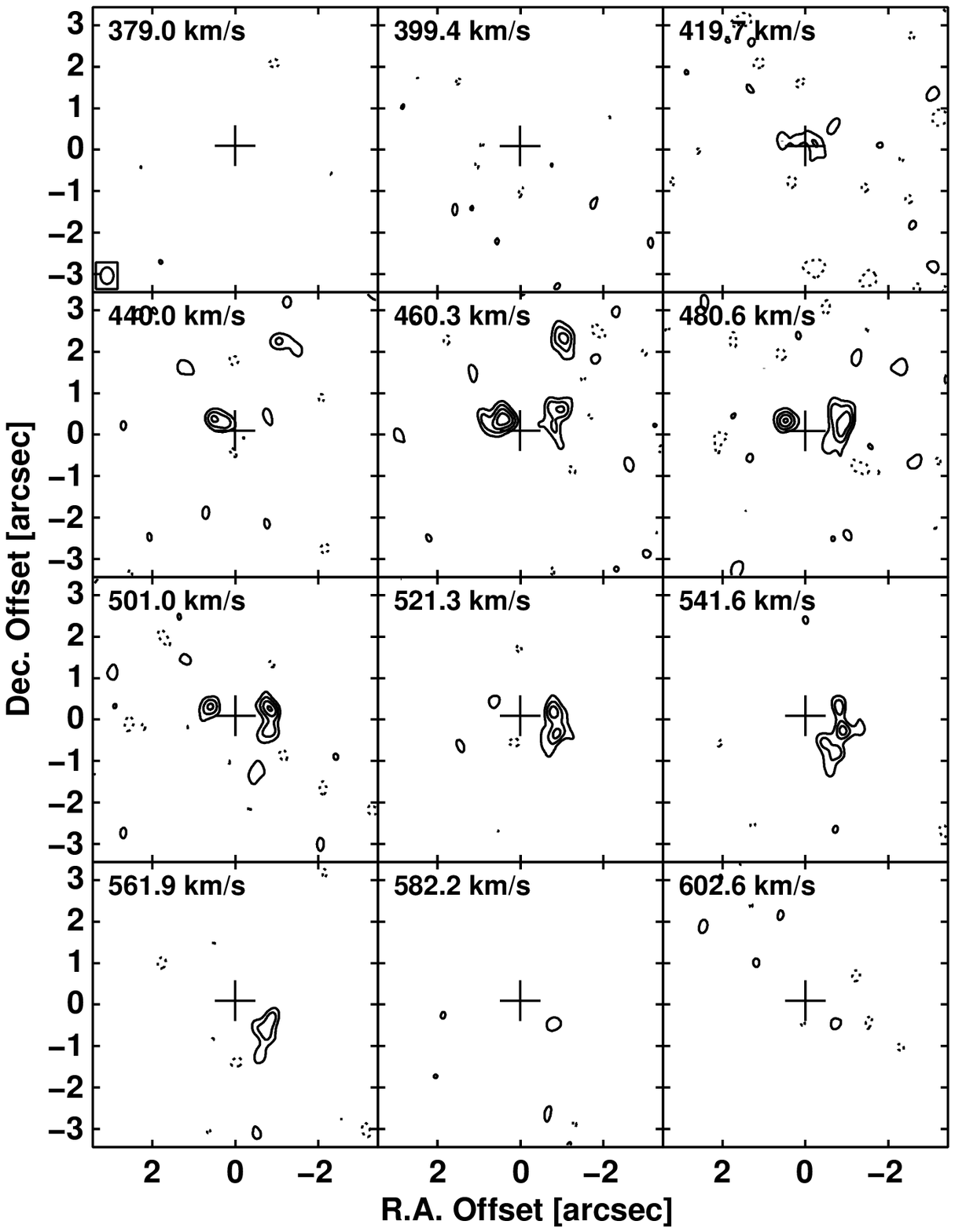}}
\caption{Channel maps of the CO(2-1) line.
	The contour levels are $-3, 3, 5, 7,$ and $9\sigma$, where
	$1\sigma$ corresponds to 5.2~mJy beam$^{-1}$ ($=0.96$~K).
	The cross in each map indicates the position of the 8.4~GHz radio
	continuum peak position of R.A.~$=13^{\rm h}29^{\rm m}52\fs7101$
	and Dec.~$=47\degr11\arcmin42\farcs696$ \citep{hag01,bra04}.
	The R.A.\ and Dec.\ offsets are the offsets from the phase
	tracking center of R.A.~$=13^{\rm h}29^{\rm m}52\fs71$ and
	Dec.~$=47\degr11\arcmin42\farcs6$.
	The synthesized beam is shown at the bottom-left corner of the
	first channel map.}
\label{fig-chan}
\end{figure}

The data were calibrated using GILDAS, and were imaged using AIPS.
The data were CLEANed with natural weighting, and the synthesized
beam sizes are $0\farcs40\times0\farcs31$ (14~pc $\times$ 11~pc) with
a position angle (P.A.) of $0\degr$ and $0\farcs85\times0\farcs55$
(29~pc $\times$ 19~pc) with a P.A.\ of $13\degr$ for CO(2-1) and
CO(1-0) images, respectively.
Fig.~\ref{fig-chan} shows the channel maps of CO(2-1) emission
with a 20.3~km s$^{-1}$ velocity resolution.
The channel maps of CO(1-0) emission show similar features to that
of CO(2-1) emission with lower spatial resolution.
Fig.~\ref{fig-mom01} shows integrated intensity and intensity
weighted mean velocity maps of the CO(2-1) and CO(1-0) lines.
The noise levels for continuum maps are 1.2~mJy beam$^{-1}$ at
1.3~mm and 0.54~mJy beam$^{-1}$ at 2.6~mm, respectively.
We did not detect any significant continuum emission at either
frequency.

\section{Results}
\label{sect-res}

\begin{figure}
\resizebox{\hsize}{!}{\includegraphics{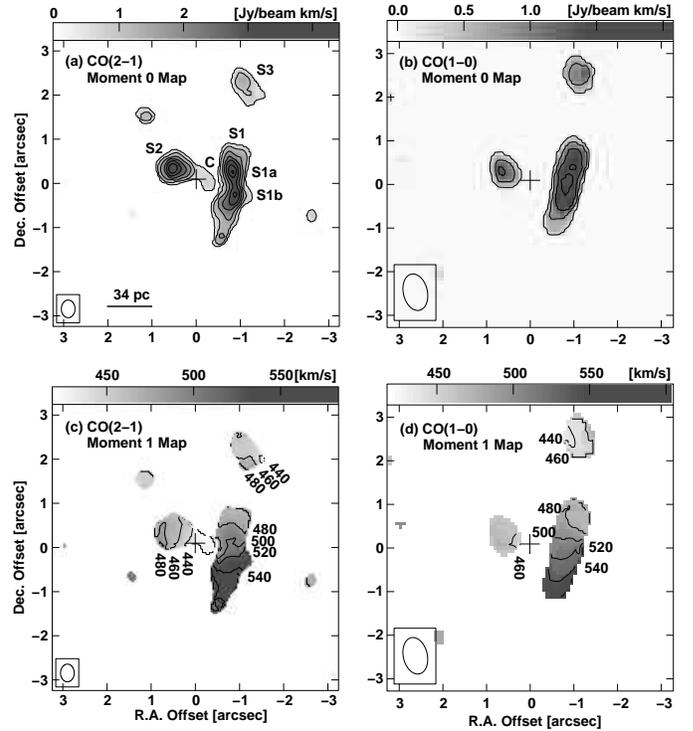}}
\caption{Integrated intensity (moment 0) and intensity weighted
	velocity (moment 1) maps of the CO(2-1) and CO(1-0) lines.
	The synthesized beams are shown at the bottom-left corner of each
	image.
	The crosses and the reference positions of the R.A.\ and
	Dec.\ offsets are the same as in Fig.~\ref{fig-chan}.
	(a) The CO(2-1) moment 0 image.
	The contour levels are (1, 3, 5, 7, 9, and 11) $\times$
	0.334~Jy beam$^{-1}$ km s$^{-1}$ ($=62.0$~K km s$^{-1}$).
	(b) The CO(1-0) moment 0 image.
	The contour levels are (1, 3, 5, and 7) $\times$
	0.257~Jy beam$^{-1}$ km s$^{-1}$ ($=50.6$~K km s$^{-1}$).
	(c) The CO(2-1) moment 1 image.
	(d) The CO(1-0) moment 1 image.
	}
\label{fig-mom01}
\end{figure}

\begin{figure}
\resizebox{\hsize}{!}{\includegraphics{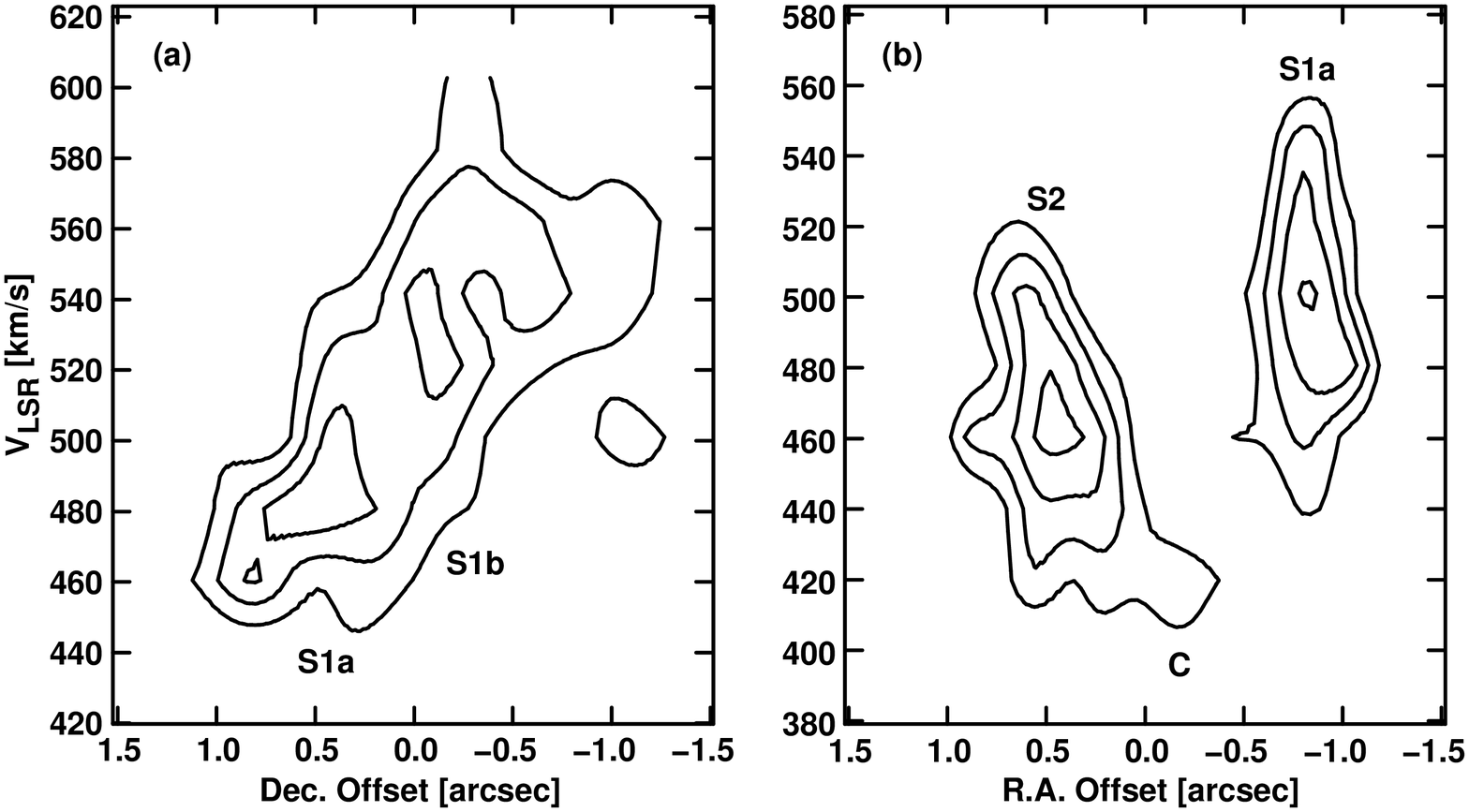}}
\caption{Position-velocity (PV) diagrams of the CO(2-1) line.
	The contour levels are $3, 5, 7,$ and $9\sigma$, where
	$1\sigma$ corresponds to 5.2~mJy beam$^{-1}$ ($=0.96$~K).
	(a) PV diagram along the north-south elongated S1 feature
		(P.A.\ of the cut is $103\degr$).
		The positions for S1a and S1b are shown with labels.
	(b) PV diagram along R.A.\ with the cut through the S1a, C,
		and S2 features (P.A.\ of the cut is $90\degr$).
		The positions for S1a, S2, and C are shown with labels.
	}
\label{fig-pv}
\end{figure}

Most of the CO(2-1) emission is detected within $\sim1\arcsec$
(34~pc) of the center, and is located mainly on the eastern and
western sides of the nucleus.
There is also weak emission located $\sim2\farcs7$ northwest of the
nucleus.
The overall distribution and kinematics are consistent with past
observations \citep{koh96,sco98}, if we degrade our image to lower
angular resolution; a blueshifted feature with the average velocity
of $\sim460$~km s$^{-1}$ at the eastern side of the nucleus, and a
redshifted feature with an average velocity of $\sim500$~km s$^{-1}$
at the western side (Figs.~\ref{fig-chan},~\ref{fig-mom01}; see also
Fig.~\ref{fig-pv}b).
We refer to these main structures with the same labels as in
\citet{sco98} (Fig.~\ref{fig-mom01}a).

Our higher resolution images, however, show more complicated
structures and kinematics than the previous low angular resolution
observations.
Molecular gas on the western side of the nucleus, S1, is elongated in
the north-south direction and separated into two main peaks
(S1a and b).
S1a is located $0\farcs9$ (30~pc) northwest of the nucleus,
and S1b is $1\farcs0$ (34~pc) to the southwest.
On the eastern side of the nucleus, the molecular gas has an
intensity peak $0\farcs6$ (20~pc) to the northeast (labeled S2),
which is located closer to the nucleus in projected distance than
that of S1a/b.

The feature S1 shows a clear velocity gradient along the north-south
direction, which is shown in Fig.~\ref{fig-mom01}(c) and also in the
position-velocity (PV) diagram (Fig.~\ref{fig-pv}a).
This gradient was previously suggested by the CO(3-2) data
\citep{mat04}, but the magnitude of the velocity gradient is
different.
The computation of the magnitude of the velocity gradient is similar
to that used for the CO(3-2) data.
The fitting result indicates a velocity gradient within S1 of
$2.2\pm0.3$ km s$^{-1}$ pc$^{-1}$, which is larger than that reported
previously, $0.77\pm0.01$ km s$^{-1}$ pc$^{-1}$ (the value has been
modified by the different distance of the galaxy used).
This difference is partially due to the larger beam size of
the previous result; the CO(3-2) data set has a beam size of
$3\farcs9\times1\farcs6$ with a P.A.\ of $146\degr$, and the
velocities of S2/C and S3 contaminate that of S1.

The CO(1-0) maps show very similar molecular gas distribution and
kinematics as those in CO(2-1) maps (Fig.~\ref{fig-mom01}b,d).
Only the western emission was detected in previous observations
\citep{aal99,sak99}, but our map clearly shows the emission from
both side of the nucleus.

In addition to the previously known features, our CO(2-1) image also
shows a weak emission near the nucleus with a structure elongated in
the northeast-southwest direction (feature C in
Fig.~\ref{fig-mom01}a).
This structure could be a part of S2, since the velocity map
(Fig.~\ref{fig-mom01}c) and the PV diagram (Fig.~\ref{fig-pv}b) show
a smooth velocity gradient, although most of the emission in C comes
from only one velocity channel
(419.7~km s$^{-1}$ map in Fig.~\ref{fig-chan}).
The velocity gradient between S2 and C is in an opposite sense to
that previously seen with the lower angular resolution observations
mentioned above.
This structure is not detected in the CO(1-0) line, but a hint of
a velocity gradient can be seen in Fig.~\ref{fig-mom01}d.

The total CO(2-1) integrated intensity of S1, S2, and C is
25.01~Jy km s$^{-1}$, and that of S1 and S2 in \citet{sco98} is
33.44~Jy km s$^{-1}$, so that our data detected $75\%$ of their
intensity.
\citet{sco98} detected $\sim50\%$ and $20\%$ of the single dish
CO(2-1) flux in redshifted and blueshifted emission, respectively, so
that our data recovered $\sim25\%$ of the single dish flux.

\section{Discussion}
\label{sect-dis}

\subsection{Jet-entrained molecular gas}
\label{sect-dis-vel}

\begin{figure}
\resizebox{\hsize}{!}{\includegraphics{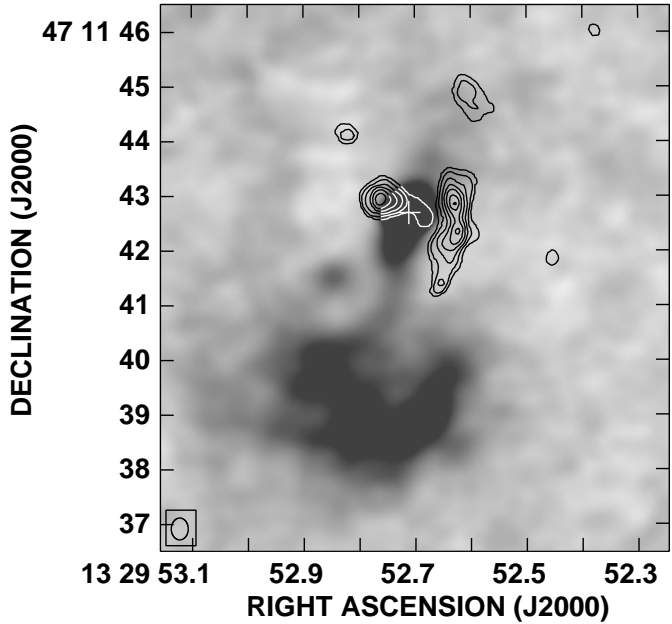}}
\caption{The CO(2-1) integrated intensity image (contours) overlaid
	on the VLA 6cm radio continuum image \citep[greyscale;][]{cra92}.
	The contour levels, the synthesized beam, and the cross are the
	same as in Fig.~\ref{fig-chan}.}
\label{fig-covla}
\end{figure}

Our molecular gas data show a clear north-south velocity gradient
within the feature S1.
We suggested from our previous study that this velocity gradient may
be due to molecular gas entrainment by the radio jet \citep{mat04}.
Here we revisit this possibility with higher spatial and velocity
resolution data.
Fig.~\ref{fig-covla} shows our CO(2-1) image overlaid on the 6~cm
radio continuum image \citep{cra92}.
The radio continuum image shows a compact radio core coincident with
the nucleus, and the southern jet emanating from there (note that
the northern jet is located outside our figure).
The CO(2-1) map clearly shows that S1 is aligned almost parallel to
the jet.
In addition, Figs.~\ref{fig-mom01} and \ref{fig-pv} show that the
velocity gradient in S1 is also almost parallel to the jet.

The velocity increases from $\sim480$~km s$^{-1}$ at S1a to
$\sim540$~km s$^{-1}$ south of S1b.
This increment is very similar to that observed in the NLR clouds
along the radio jet;
\citet{bra04} measured the velocities and velocity dispersions of the
clouds using the [\ion{O}{III}] $\lambda5007$ line, and showed that
the velocity of the southern $\lse1\arcsec$ clouds from the nucleus
are at V$_{\rm LSR}\sim440-590$~km s$^{-1}$ and the velocity
increases as the clouds move away from the nucleus (see Table 2 and
Fig.~9 of their paper)\footnote{We selected the clouds with a
velocity dispersion of less than 100~km s$^{-1}$; Clouds 3, 4,
and 4a in \citet{bra04}.
If we include all the clouds, the velocity is
$\sim440-690$~km s$^{-1}$ with a range of velocity dispersion of
$\sim25-331$~km s$^{-1}$; Clouds 2, 3, 3a, 4, 4a, and 4b.}.
This velocity range and increment are consistent with our data.
Furthermore recent observations of H$_{2}$O masers toward the nucleus
also show a velocity gradient along the jet with the same sense as
our results \citep{hag07}, in addtion to the good correspondance of
the velocity range \citep{hag01,hag07,mat04}.
These results suggest that the molecular gas in S1 (and the NLR
clouds and the H$_{2}$O masers) is possibly entrained by the radio
jet.
These results also suggest that some of the material in NLRs is
supplied from molecular gas close to AGNs.

Another example of jet-entrained neutral gas is found in the radio
galaxy 3C293 \citep{emo05}.
The velocity of \ion{H}{I} gas in absorption spectra toward the AGN
matches that of ionized gas along kpc-scale radio jets.
The spatial coincidence is not clear, since the spatial resolution of
the \ion{H}{I} data is lower ($25\farcs3\times11\farcs9$) than that
of the ionized gas data.
Our result is therefore the first possible case of entrainment of
molecular gas by a jet at the scale of ten pc.

The better resolution of our new CO data allows us to revisit the
values of the molecular gas mass, momentum, and energy of the
entrained gas.
We derive $6\times10^{5}$~M$_\odot$, $8\times10^{45}$~g cm s$^{-1}$,
and $3\times10^{52}$ ergs for these quantities.
These values are about half of the previous values derived from the
CO(3-2) data, mainly due to the larger beam, but the conclusion is
similar;
the energy of the entrained gas could be similar to that of the radio
jet \citep[$>6.9\times10^{51}$~ergs;][]{cra92}, but the momentum is
much larger than that of the jet ($2\times10^{41}$~g cm s$^{-1}$).
One way to explain this discrepancy is through a continuous input of
momentum from the jet
\citep[see][for more detail discussions]{mat04}.

\subsection{Obscuring material around the Seyfert 2 nucleus}
\label{sect-dis-obs}

The feature C is located in front of the Seyfert 2 nucleus, and the
CO(2-1) intensity is about 62.0~K km s$^{-1}$ (Fig.~\ref{fig-mom01}a).
Hence the column density can be calculated as
$6.2\times10^{21}$~cm$^{-2}$ using a CO-to-H$_{2}$ conversion
factor of $1.0\times10^{20}$~cm$^{-2}$ (K km s$^{-1}$)$^{-1}$
\citep{mat04} and assuming a CO(2-1)/(1-0) ratio of unity.
This value is far lower than that derived from the X-ray absorption
of $5.6\times10^{24}$~cm$^{-2}$ \citep{fuk01}.
As is mentioned in Sect.\ref{sect-res}, the missing flux of our data
is $\sim75\%$.
However, even if all of this missing flux contributes to obscuring
the nuclear emission, this large column density difference cannot be
explained.
Changing the conversion factor or the ratio by an order of magnitude
also cannot explain this large difference.
One way to reconcile this disparity is to assume that C is not
spatially resovled, in which case the computed column density is a
lower limit.
Alternatively, the obscuring material preferentially traced by
higher-J CO lines or denser molecular gas tracers such as HCN may be
involved.
The CO(3-2) intensity in brightness temperature scale is $\sim2$
times stronger than that of CO(1-0) \citep{mat04}, and the HCN(1-0)
intensity is also relatively stronger
\citep[HCN/CO~$\sim0.4$;][]{koh96} than normal galaxies.

\subsection{Molecular gas at ten pc scale from the Seyfert nucleus}
\label{sect-dis-nuc}

Previous studies suggest that the blue shifted eastern feature S2 and
the red shifted western feature S1 may be the outer part of a
rotating disk as in the AGN unified model.
However, our images show a more complicated nature, and no clear
evidence of simple disk characteristics.

The simplest interpretation is that S1 and S2/C are independent
structures.
Since S1 is affected by the jet but S2/C is not, S1 is expected to be
located closer to the nucleus than S2/C, and the projection effect
makes the position of S2/C closer in our images.
Alternatively, S2/C may be close to the nucleus, but the entrained
gas has been already swept away or ionized by the jet.
S2/C has a velocity gradient, and therefore can be interpreted as a
streaming gas, presumably infalling toward the nucleus, as is
observed in the Galactic Center \citep{lo83,ho91}.

S1 and S2/C can also be interpreted as a rotating disk that is
largely disturbed by the jet, and only a part remains.
According to the velocity gradient along S2/C, the blueshfited gas is
expected at S1, which is the opposite sense to the previous
suggestion, but the gas shows no signs of it due to the jet
entrainment.
This is possible from the timescale point of view; under this
interpretation, S1 should have a blueshifted rotation velocity of
$\sim380$~km s$^{-1}$ based on the velocity gradient in S2/C.
S1 has a velocity $\sim150$~km s$^{-1}$ higher than the expected
rotational velocity, and we assume that this is the entrained
velocity.
In this case, it takes $2\times10^5$ years to be elongated along the
jet by $\sim1\arcsec$ or 34~pc.
On the other hand, the rotation timescale at this radius is about
$2\times10^6$ years, an order of magnitude longer timescale.
The rotating disk can therefore be locally disturbed by the jet.

However, the above two explanations have difficulty in explaining
optical images of the nucleus; the Hubble Space Telescope images show
``X'' shaped dark lanes in front of the nucleus \citep{gri97},
suggesting the existence of a warped disk or two rings with one
tilted far from another.
An alternative explanation of the dark lanes is that, as previously
proposed, there is a rotating edge-on ring with S2 as blueshifted gas
and S1a as redshifted gas.
In this case, the feature C can be the counterpart of another dark
lane, which runs northeast-southwest, although C has to be a
counter-rotating or Keplarian rotating disk to explain the opposite
sense of the velocity gradient to that of the S1a/S2
(Sect.~\ref{sect-res}).
This configuration explains the ``X'' shape, but has a rather
complicated configuration, and it is difficult to explain why the
inner disk C is not disturbed by the jet.

\hspace*{1ex}

We imaged the nuclear region of the Seyfert 2 galaxy M51 at
$\sim10$~pc resolution, and we see no clear evidence of a
circumnuclear rotating molecular gas disk as previously suggested.
The molecular gas along the radio jet is most likely entrained by
the jet.
The explanations for other gas components are speculative, possibly
involving a circumnuclear rotating disk or streaming gas.

\begin{acknowledgements}
We thank Arancha Castro-Carrizo and the IRAM staff for the new A
configuration observations.
We also thank the anonymous referee for helpful comments.
IRAM is supported by INSU/CNRS (France), MPG (Germany) and IGN
(Spain).
This work is supported by the National Science Council (NSC) of
Taiwan, NSC 95-2112-M-001-023.
\end{acknowledgements}

\end{document}